\documentclass[prb,amsmath,amssymb]{revtex4}% Physical Review B

\usepackage{graphicx}% Include figure files
\usepackage{dcolumn}% Align table columns on decimal point
\usepackage{bm}% bold math

\begin{document}

%\preprint{SMU/01-05}

\title{A Compact Apparatus for Muon Lifetime Measurement and\\ Time
Dilation Demonstration in the Undergraduate Laboratory}

\author{Thomas Coan}
\email{coan@mail.physics.smu.edu}
\author{Tiankuan Liu}
\email{liu@mail.physics.smu.edu}
\author{Jingbo Ye}%
 \email{yejb@mail.physics.smu.edu}
\affiliation{Physics Department, Southern Methodist University, Dallas, TX 75275 USA}%


\begin{abstract}
We describe a compact apparatus that automatically measures the charge
averaged lifetime of atmospheric muons in plastic scintillator using
low-cost, low-power electronics and that measures the stopping rate of
atmospheric muons as a function of altitude to demonstrate
relativistic time dilation.  The apparatus is designed for the
advanced undergraduate physics laboratory and is suitable for field
measurements.
\end{abstract}

\pacs{Valid PACS appear here}% PACS, the Physics and Astronomy
\maketitle

\section{\label{sec:intro}Introduction}

Measurement of the mean lifetime of muons produced in
Earth's atmosphere from collisions between cosmic rays and air nuclei
is a common experiment\cite{hist1,hist2,hist3} in advanced
undergraduate physics laboratories.  Typically, a single scintillating
medium, massive enough to stop some useful fraction of the muons
impinging on it, is viewed by one or two photomultiplier tubes (PMTs)
that detect the pair of scintillation light flashes associated with
the entry and subsequent decay of a stopped muon. Histogramming the
time interval between the two flashes and then fitting the time
distribution with an exponential function yield the mean muon
lifetime. Various PMT readout and histogramming techniques have been
implemented to produce the decay time histogram. However, such
techniques tend to rely on relatively expensive and bulky electronic
instrumentation (e.g., NIM and CAMAC-standard modules) and stand-alone
multi-channel analyzers to generate the decay time histogram. We have
developed fully equivalent readout instrumentation, based on a complex
programmable logic device (CPLD\cite{cpld}), that is compact ($20\times
25\times5\,{\rm cm^3}$), low cost and low power ($<25\,$W).  The
readout instrumentation is easily interfaced to a laptop computer
to display and fit the decay time histogram.

\section{\label{sec:det} Detector and Readout Electronics}

We use a standard detector configuration, with no attempt made to
select only vertically traveling muons. A plastic scintillator in the
shape of a right circular cylinder ($15.2\,{\rm cm}$ diameter and
$12.7\,{\rm cm}$ tall) is viewed by a single 10-stage, $51\,$mm
diameter, bi-alkali photocathode PMT biased to nominal gain $3\times
10^5$ attached to one end. Both scintillator and PMT are wrapped
carefully with aluminum foil and electrical tape to prevent light
leaks. The PMT is biased using a compact, commercially available DC-DC
converter\cite{emco} with negative high voltage (HV) applied to its
photocathode. To mimic events where a muon enters, stops and then
decays inside the scintillator, a blue light emitting diode (LED) is
permanently inserted into a small hole drilled in one end of the
scintillator. The LED can be driven by a transistorized pulser circuit
that produces pairs of pulses at a nominal repetition rate of 100 Hz
with an adjustable interpulse separation in a given pair from
$300\,{\rm ns}$ to $30\,\mu{\rm sec}$. For robustness and portability,
and so that no HV electrodes are exposed to students, the
scintillator, PMT, HV circuit and pulser are all enclosed inside a
black anodized aluminum tube $36\,$cm tall and $15.2\,\text{cm}$ inner
diameter. The cylinder is capped at both ends.  Power to the HV supply
and pulser circuitry is provided by a single multi-connector cable and
the PMT signal is sent to the readout electronics module by a coaxial
cable. Potentiometers installed in the cylinder cap allow student
adjustment of the PMT HV and LED interpulse time separation. Provision
is made to monitor the HV with a conventional voltmeter and output of
the pulser circuitry is accessible by a coaxial connector on the cap.
The mass of the overall detector is $5\,$kg.

\begin{figure}[ht]
%\rule{5cm}{0.2mm}\hfill\rule{5cm}{0.2mm}
%\vskip 2.5cm
%\rule{5cm}{0.2mm}\hfill\rule{5cm}{0.2mm}
%\hspace{3.5cm} 
\includegraphics[scale=.50]{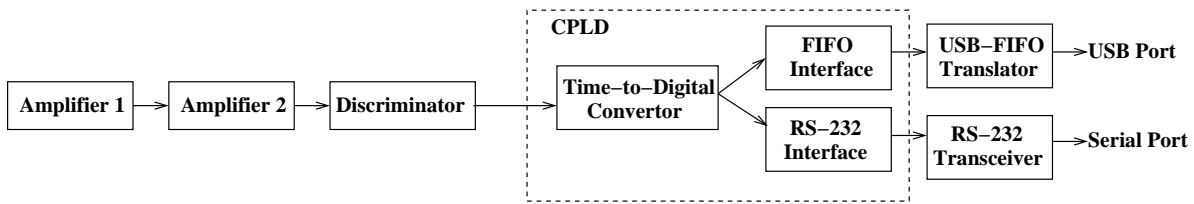}
\caption{Block diagram of the readout electronics showing the
two-stage amplifier, discriminator, CPLD and I/O communications
circuitry. Signals are processed from left to right.
\label{fig:box}}
\end{figure}

The electronic circuitry to process PMT signals, perform timing,
communicate with the computer, and provide power to the detector is
mounted on one printed circuit board (PCB) located inside a single
enclosure of volume $20\times 25\times5\,{\rm cm^3}$.  A block diagram
of this circuitry is shown in Fig.~\ref{fig:box}.  Signals from the
PMT anode are coupled by the coaxial cable to the input of a two-stage
amplifier constructed from a fast current feedback
amplifier\cite{amp}. A typical raw PMT signal amplitude for muon decay
events is $100\, $mV into $50\,\Omega$ impedance. The amplifier output
feeds a discriminator with adjustable threshold and TTL logic
output. Students can monitor the amplifier and discriminator outputs
via BNC connectors mounted on the front of the electronics enclosure.

The discriminator output signal is processed by a CPLD, mounted on a
single PCB. A CPLD is a single, flexible integrated circuit (IC)
comprised of thousands of usable logic gates that can implement a wide
variety of digital logic functions. The chip includes programmable
interconnections and multiple input/output (I/O) pins, and can be
clocked at rates up to $\sim 100\,$MHz. Its behavior can be
extensively and reliably simulated before its controlling program is
downloaded into its electrically erasable programmable read only
memeory (EEPROM). Such ICs are reprogrammable and relatively cheap,
typically costing a few tens of dollars, making the inexpensive
implementation of a wide variety of digital logic signal processing
circuitry practical.

The CPLD is programmed in a digital hardware description language
(VHDL) to function primarily as a timer. A logical ``yes'' output from
the discriminator, corresponding to an amplified PMT signal above
threshold, causes either the CPLD to start recording clock cycles from
the 50 MHz crystal oscillator that clocks it, or, if the CPLD had
already started counting clock cycles, to stop counting.  In this way,
any amplified PMT signal above threshold can serve as a stop or start
timing signal. The CPLD is also programmed to reset itself if a second
``yes'' does not occur within $20\,\mu$sec of the first one. This
simple logic scheme corresponds to our desired event scenario where
we have a flash of scintillator light corresponding to a muon entering
and stopping in the scintillator, and a second flash occurring when
the stopped muon decays.

The CPLD formats data in a simple fashion. For successive PMT signals
above threshold and within the $20\,\mu$s timing window, the time
difference, in integral units of $50\,$MHz clock cycles, between these
two signals is recorded. Data of this type is ultimately histogrammed
and fit to an exponential curve by the laptop software to extract the
muon lifetime. For cases when there are no signal pairs within the
timing window, the CPLD merely records the integer 1000, corresponding
to the number of clock cycles in our timing window. All data are
subsequently sent to the laptop through either a serial or USB port.

The CPLD I/O circuitry (see Fig~\ref{fig:box}) has two physical ports
to simplify interfacing to laptop computers. One port is a standard
serial port that follows the RS-232 protocol and that relies on a
dedicated RS-232 transceiver chip to shift RS-232 standard voltage
levels to low voltage TTL levels to communicate with the RS-232
interface module resident in the CPLD. The data transmission rate
between CPLD and laptop is 115~kbits/s.  The other port adheres to the USB
1.1 protocol and relies on a USB-FIFO (``first in, first out'')
translator chip to communicate with the FIFO interface module within
the CPLD. Data transmission rates in this case are 2.4~Mbits/s.

Overall power consumption of the electronics module is less than
$25\,$W, sufficiently low that it can be powered in the field from an
automobile cigarette lighter.

\section{Data Display Software}

The laptop-resident software that displays and curve fits the decay
time histogram is written in the Tcl/Tk scripting language, an open
source language that permits easy implementation of graphical user
interfaces and that is compatible with Unix, Microsoft Windows and
Apple Macintosh operating systems. The laptop continuously examines
its own I/O port buffers for the presence of any data that the CPLD
has sent it.  Any data consistent with PMT pulse pairs within the
timing window has the corresponding pulse separation time entered into
a decay time histogram. Data not corresponding to pulse pairs is used
to update various rate meters monitoring the frequency of PMT signals
above threshold.  All data is then slightly reformatted to include the
absolute time in seconds when it was examined by the laptop before
being written to disk in ASCII format for easy human interpretation
and exporting to student written data analysis software routines. The
histogram and rate meters are displayed in real time for student
observation.

The laptop software has provision for simulating muon decay by randomly
generating times according to an exponential distribution with a user
selectable lifetime. This permits students to practice their curve
fitting and lifetime extracting software routines on large simulated
data sets.

\section{Mean Muon Lifetime}

\begin{figure}[h]
%\rule{5cm}{0.2mm}\hfill\rule{5cm}{0.2mm}
%\vskip 2.5cm
%\rule{5cm}{0.2mm}\hfill\rule{5cm}{0.2mm}

\includegraphics[scale=0.5]{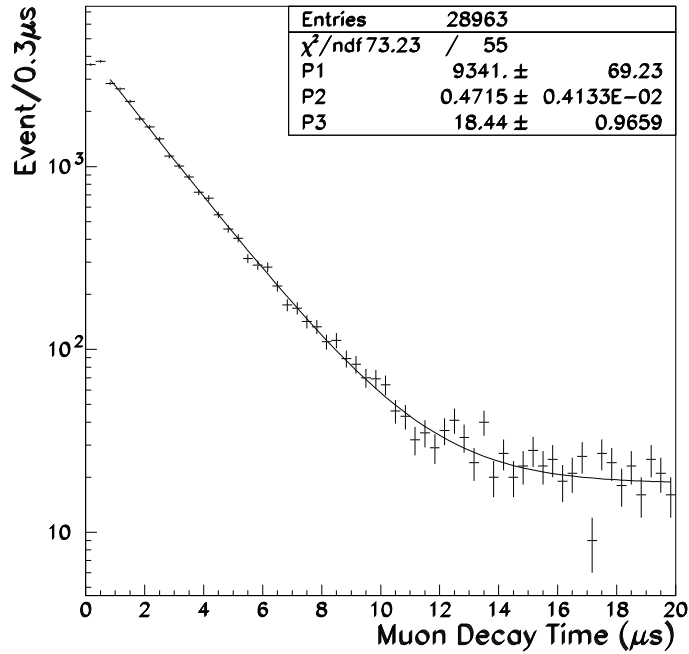}
\caption{Decay time histogram for 28,963 events collected over 480
hrs. The dots with error bars are data and the line is a fit to the
data with the three-parameter function $f(t)=
P_1P_2\exp(-P_2t)+P_3$. The values for the fit parameters plus the
$\chi^2$ per degree-of-freedom are shown in the upper right hand
corner.
\label{fig:histo}}
\end{figure}

A decay time histogram for muons stopping in our detector formed by
histogramming the time between two successive scintillator flashes
within our $20\,\mu$s timing window is shown in Fig.~\ref{fig:histo}.
Dots with crosses are data and the line is a fit to the data. This
histogram contains $28,963$ events collected over 480 hours of running
and contains $\mu^+\text{and}\ \mu^-$ decays as well as
background. The data is fit to the functional form
$\dot{N}(t)=P_1P_2\exp{(-P_2t)} + P_3$, characteristic of radioactive
decay with background. Here $\dot{N}(t)$ represents the observed
number of decays per unit time at time $t$ and the quantities $P_1,
P_2^{-1} \text{and\ } P_3$ are constants extracted from the fit and
represent an overall normalization constant, the muon lifetime and the
background level, respectively.  These values are shown in the box in
the upper right-hand corner of Fig.~\ref{fig:histo}. The quality of
the fit is indicated by the low $\chi^2$ per degree-of-freedom
($\chi^2/\text{d.o.f.}= 73/55$). The fit was done with PAW, a freely
available\cite{paw} fitting and plotting software package. Other
fitting packages return similar results. Due to the properties of an
exponential function, it is irrelevant that the muons whose decays we
observe are undetected when born outside the detector.

The extracted value of the mean muon lifetime $\tau= P_2^{-1}
=2.12\pm0.02\,\mu$s (statistical error only) is less than the free
space value $\tau_{\mu}= 2.19703\pm 0.00004\,\mu$s due to the
non-negligible probability that a $\mu^-$, but not a $\mu^+$, will be
captured into the K-shell of a scintillator carbon atom and then be
absorbed by its nucleus\cite{wardetal}. (The probability that a
stopped $\mu^-$ will be absorbed by a target atom of atomic number $Z$
is proportional to $Z^4$.)

The extracted background rate of fake muon decays is inferred from the
value of $P_3$ and the 480 hr running time, and corresponds to
$0.6\,$mHz.  This rate of two PMT signals in coincidence is consistent
with the observed rate of single PMT signals above threshold ($\sim
6\,$Hz) and our $20\,\mu$s timing window.  For comparison, the
fitted rate of muon decays in the scintillator is $17\,$mHz ($\sim
1\,\text{min}^{-1}$).

From the charge averaged lifetime of muons in what is essentially a
carbon target (the ratio of hydrogen to carbon in plastic scintillator
is ~1:1), and the lifetime of $\mu^-$ in carbon\cite{reiter}, it is
straightforward\cite{rossi} to measure the charge ratio abundance
$\rho=N(\mu^+)/N(\mu^-)$ of low-energy ($E\lesssim 200\,$MeV) muons at
sea-level. For example, from our measured lifetime, we find $\rho=
1.08\pm0.01$ (statistical error only), averaged over the angular
acceptance of the detector, a value consistent with the diminishing
trend\cite{inpc} for $\rho$ as the muon momentum approaches zero.

\section{Demonstration of Relativistic Time Dilation}

The stopping rate of muons in the detector as a function of altitude
above sea level can be used to demonstrate relativistic time dilation.
Although the detector design is non-optimal for this demonstration
since it is sensitive to muons with a range of velocities as well as
non-vertical trajectories, it has the advantage that no bulky velocity
selecting absorbers or additional trajectory defining
scintillators\cite{easwar} are required. The idea is simple. The total
number of stopped muons in the detector in some fixed time interval
and at some fixed altitude above sea level (a.s.l.) is measured from
the decay time time histogram.  A lower altitude is selected and
predictions made for the new stopping rate that do and do not include
the time dilation effect of special relativity. Measurement then
discriminates between the two predictions.

To make the comparison between the competing assumptions meaningful,
the effects of the energy loss of a muon as it descends in the
atmosphere as well as the shape of the sub-GeV/c muon momentum
spectrum\cite{muspec} near sea-level should be included. The first
effect tends to increase the transit time of the muon from one
altitude to another and the other tends to over emphasize the
effects of time dilation.

The transit time $t^{\prime}$ measured in the muon's rest frame as it
descends vertically in the atmosphere from a height $H$ down to
sea-level is given by

\begin{equation}
t^{\prime}=\int^0_H {\frac {dh}{c\beta(h)\gamma(h)}}\label{time}
\end{equation}

\noindent where $\beta$ and $\gamma$ have their normal relativistic
meanings, $dh$ is a differential element of pathlength and $c$ is the
speed of light. All quantities on the right-hand side of
Eq.~(\ref{time}) are measured in the detector rest frame. As the muon
descends it loses energy in a manner described by the Bethe-Bloch
equation\cite{leo} so the integral can be evaluated numerically if
great precision is desired. Instead, we use the common approximation
that a singly-charged relativistic particle loses energy by ionization
and excitation of the medium it traverses with a magnitude $dE/dx=
2\,{\text {MeV}\,\text{g}^{-1}\text{cm}{}^2}\,(\equiv\! S_0)$.
Eq.~(\ref{time}) becomes

\begin{equation}
t^{\prime} \simeq {\frac{mc} {\rho S_0}}\int^{\gamma_1}_{\gamma_2}{\frac{d\gamma}
{\sqrt{\gamma^2 -1}}}\label{time2}
\end{equation}

\noindent Here, $\gamma_1$ is the muon's Lorentz factor at height $H$,
$\gamma_2$ is its Lorentz factor just before it enters the sea level
scintillator, $m$ is the muon mass and $\rho$ denotes the
pathlength-averaged mass density of the atmosphere.  We take
$\gamma_2\simeq 1.5$ since we want muons that stop in the scintillator
and assume, consistent with our detector geometry, that stopped muons
travel an average distance $s=10\, \text {g/cm}^2$ in the scintillator
. (See muon range-momentum graphs from the Particle Data
Group\cite{pdg} for correlating a muon's range with its momentum.) The
appropriate value of $\gamma_1$ depends on the height $H$ where we
take our upper measurement and is computed from the energy $E_1$ a
muon has at that height if it is to arrive at the sea-level detector
with $\gamma_2=1.5$ (corresponding to energy $E_2=160\,$MeV). Clearly,
if a muon loses energy $\Delta E$ in traversing a vertical distance
$H$, then $E_1 = \Delta E + 160\,$MeV. The quantity $\Delta E$ can be
computed from the Bethe-Bloch equation or estimated from the above
rule-of-thumb for minimum ionizing particles and properties of the
standard atmosphere.

Since the time dilation demonstration relies on stopping muons in the
detector, we must account for the fact that muons that eventually stop
in the lower detector have, at the position of the upper detector, an
energy that is greater than those muons that would be stopped in the
upper detector. Since the momentum spectrum of sub-GeV muons near
sea-level is not flat, but peaks at muon momentum
$p_{\mu} \sim 500\,$MeV/c, we correct for this affect so that the
effective flux of incident muons is appropriately normalized. (This is
easy to see if you assume muons don't decay at all and only lose
energy in the atmosphere as they descend.) We do this by measuring the
ratio of stopping rates at a pair of altitudes to determine a single
scaling factor that we can apply to other pairs of altitudes.

To illustrate the procedure, we measure the muon stopping rate at two
different elevations ($\Delta h=3,000\,$m between Taos, NM and Dallas,
TX) and compute the ratio $\text R_{\text {obs}}$ of observed stopping
rates ($\text R_{\text{obs}}(\text {Dallas}/\text{Taos})=
0.41\pm0.05$.)  The transit time $t^{\prime}$ in the muon's rest frame
for vertical trajectories between the two elevations is computed using
Eq.~\ref{time2} and yields $t^{\prime}=1.32\tau_{\mu}$. This
corresponds to a naive theoretical stopping rate ratio
$R=\exp(-t^{\prime}/\tau_{\mu})=0.267$. The double ratio $R_0 =
R_{\text{obs}}/R =1.5\pm 0.2$ is then interpreted as a correction
factor for the shape of the muon momentum spectrum. Note that this
same correction tends to account for muons with non-vertical
trajectories that stop in the detector. These slanting muons with a
projection onto the vertical axis of distance $H$ travel further in
the atmosphere and hence start with more energy than their purely
vertical counterparts.

To verify that the procedure is sensible, we choose a new elevation
($H=2133\,$m a.s.l. at Los Alamos, NM), compute the transit time
$t^{\prime}=1.06\tau_{\mu}$ and the expected stopping rate ratio
(Dallas/Los Alamos) $R_{\text {thy}} =
R_0\exp(-t^{\prime}/\tau_{\mu})=0.52\pm 0.06$. The observed ratio
$R_{\text obs}=0.56\pm0.01$, showing good agreement.
Table~\ref{tab:stopdata} summarizes relevant measurements and lists in
the third column calculated proper transit times for vertical muon
trajectories, in units of the proper muon lifetime $\tau_\mu$ and
relative to Dallas' elevation.

\begin{table}
\caption{\label{tab:stopdata}Muon Stopping Rate Measurements and
Calculated Proper Transit Times}
\begin{ruledtabular}
\begin{tabular}{lcc}
Elevation (meters a.s.l.)& Observed Stopping Rate (min${}^{-1}$)& Proper transit time ($\tau_{\mu}$)\\
146   & $1.24\pm0.01$ & $-$\\
2133  & $2.21\pm0.05$ & $1.06$\\ 
3154  & $3.00\pm0.34$ & $1.32$\\
\end{tabular}
\end{ruledtabular}
\end{table}

To compare the stopping rate measurements with the competing
assumption that there is no time dilation effect (``ntd''), we proceed
as before except we calculate the transit time in the detector rest
frame and we assume all muons travel at the speed of light so as to
{\it minimize} the effect of time dilation. For the case of transit
between Los Alamos and Dallas, the transit time $t_{\text {ntd}}$ in
the detector rest frame is $t_{\text {ntd}}=6.62\,\mu$s, implying an
expected stopping rate ratio $R_{\text {ntd}}=R_0\exp(-t/\tau)
=0.08\pm 0.01$, a result strongly disfavored by observation.

\section{Summary}

We have designed a compact and low-cost apparatus for measuring the
mean muon lifetime and for demonstrating relativistic time dilation
suitable for undergraduate teaching. An electronics schematic and
Tcl/Tk data acquisition/display software are available upon request.

\begin{acknowledgments}
The technical assistance of H. van Hecke and L. Lu is greatly
appreciated, as well as the financial support of the Lightner-Sams
foundation.
\end{acknowledgments}

%\subsection{\label{sec:level2}Second-level heading: Formatting}

%\subsubsection{\label{sec:level3}Third-level heading: References and Footnotes}

\bibliography{muon}% Produces the bibliography via BibTeX.

\end{document}